\documentclass[referee]{raa}           
\usepackage{graphicx,times,subcaption}
\usepackage{natbib}
\usepackage{amssymb,amsmath}
\usepackage{multirow}
\usepackage{epstopdf}
\bibpunct{(}{)}{;}{a}{}{,}
\newcommand{\minitab}[2][l]{\begin{tabular}{#1}#2\end{tabular}}

\usepackage[a4paper=true,pagebackref=true]{hyperref}
\hypersetup{pdftitle = A Sodium laser guide star coupling efficiency measurement method, pdfauthor = Feng Lu, pdfsubject= Sodium laser guide star, pdfkeywords = sodium laser guide star adaptive optics} 
\hypersetup{colorlinks = true, linkcolor = green, anchorcolor = red, citecolor = blue, filecolor = red, pagecolor = red, urlcolor = red}

\begin{document}

   \title{A Sodium laser guide star coupling efficiency measurement method
\footnotetext{\small $*$ Supported by the National Natural Science Foundation of China, Granting Number 11303056.}
}

 \volnopage{ {\bf 2016} Vol.\ {\bf X} No. {\bf XX}, 000--000}
   \setcounter{page}{1}

   \author{Feng Lu\inst{1}, Zhi-Xia Shen\inst{1}, Suijian Xue\inst{1}, Yang-Peng Li\inst{1},  Kai Jin\inst{2}, Angel Otarola\inst{4}, Yong Bo\inst{3}, Jun-Wei Zuo\inst{3}, Qi Bian\inst{3}, Kai Wei\inst{2}, Jing-Yao Hu\inst{1}}

   \institute{Key Laboratoryb of Optical Astronomy, National Astronomical Observatories, Chinese Academy of Sciences, Beijing 100012, China; {\it jacobfeng@bao.ac.cn}\\
        \and
             The Institute of Optics and Electronics, Chinese Academy of Sciences,
             Chengdu 610209, China\\
		\and
			Technical Institute of Physics and Chemistry, Chinese Academy of Sciences, Beijing 100190, China\\
		\and
		Thirty Meter Telescope Corporation, Pasadena, California, United States\\
		\vs \no
	   {\small Received 2016; accepted 2016}
}

\abstract{Large telescope's adaptive optics (AO) system requires one or several bright artificial laser guide stars to improve its sky coverage. The recent advent of high power sodium laser is perfect for such application. However, besides the output power, other parameters of the laser also have significant impact on the brightness of the generated sodium laser guide star mostly in non-linear relationships. When tuning and optimizing these parameters it is necessary to tune based on a laser guide star generation performance metric. Although return photon flux is widely used, variability of atmosphere and sodium layer make it difficult to compare from site to site even within short time period for the same site. A new metric, coupling efficiency is adopted in our field tests. In this paper, we will introduce our method for measuring the coupling efficiency of a 20W class pulse sodium laser for AO application during field tests that were conducted during 2013-2015.
\keywords{instrumentation: adaptive optics; methods: observational; atmospheric effects}
}

   \authorrunning{L. Feng et al. }            
   \titlerunning{A Sodium laser guide star coupling efficiency measurement method}  
   \maketitle

%
\section{Introduction}           
\label{sect:intro}

Adaptive optics is one of the latest technology that significantly improved the performance of large ground-based astronomical telescope in terms of image sharpness and sensitivity. The functioning of the system relies strongly on the detection performance for the turbulence induced abberated wavefront which coming from a bright on-sky reference source that should be within isotropic angle from the observed target (\cite{1998aoat.book.....H}). The sky coverage of such bright stars is reported to be less than 1\% in the near infrared band (\cite{10.1086/316120}), which severely limits the application of AO system. The introduction of artificial laser guide star technology alleviates this problem. By projecting a suitable format laser in close direction of the observed target, one could generate an artificial guide star in the sky that will lowers the requirement of the brightness of Natural Guide Star (NGS), thus improves the sky coverage of the AO system. There are two methods in generating laser guide star, one is taking advantage of Rayleigh backscatter induced by light scattered by large molecules and dust in the lower atmosphere (0$\sim$20km), another is by exciting the sodium atoms in high atmosphere (90$\sim$110km) with sodium laser and using the resonant fluorescence of sodium atom as the reference signal. Because the sodium laser guide star has a higher altitude which is beneficial for sensing a larger volume of turbulence than Rayleigh laser guide star, it is preferable for laser guide star generation.

The brightness of the sodium laser guide star has direct impact on the wavefront detection performance of adaptive optics system. Pulsed laser when combined with range-gating technique, could avoid the “fratricide effect” which is caused by the Rayleigh backscatter in lower atmosphere. The first generation sodium pulse lasers have output power merely of a few-watts level. However, it is soon found out that by further increasing the output power or reducing the pulse width, the returned flux would be easily saturated. Theoretical modeling (\cite{Holzloehner2010}, \cite{2010SPIE.7736E..0VH}, \cite{2012SPIE.8447E..4LR}) shows that it is necessary to tune laser's temporal/spectrum behavior as well as other characteristic parameters to take advantange of the physics of sodium atom to further increase the returned flux. Optimum laser's parameter set has to be determined with on-sky test based on certain metric which should be able to reflect the absolute performance of the laser during laser guide star generation. In earlier papers, the metric used was often reported to be returned photon flux measured by differential photometry with Johnson V band filter. This metric is fine for tuning if the duration of the test is short comparing to the variability of sodium abundance. However, the sodium abundance in the atmosphere is possible to change from $2\times10^{13}$ to more than $10\times10^{13} \text{ atoms/m}^2$ in one night (\cite{Pfrommer2009}), and it could be even higher in short term due to sporadic pocket of sodium concentration in the atompshere. A new metric was used in Holzloehner's simulation paper (\cite{Holzloehner2010}), coupling efficiency of the sodium laser, which was formerly used in Lidar equation. We repeat the Lidar equation hereby in equation \ref{eq:lidar equation}. The coupling efficiency of the laser $s_{ce}$ is on the left side. On the right side, the returned flux in unit receiver area $F$ (unit photons/s/m$^2$) is normalized with laser power in mesosphere $P(T_a)^X$, the sodium column abundance $C_{Na}$ and considering the airmass $X$ and the height of the sodium layer $L$. 

\begin{equation}
\label{eq:lidar equation}
s_{ce} = \frac{F L^2}{P (T_a)^{2X} C_{Na} X}
\end{equation}

The coupling efficiency thus has the advantage that it is invariant from changes in sodium abundance, sodium layer height, atmospheric transparency, laser power variations if all parameters in the equation could be measured synchronously at the same location. The complexity and cost of the measurement system hinders the popularity of this metric. However, because it directly reflects the absolute performance of the laser in generating laser guide star, it is the most helpful metric for optimizing sodium laser's internal parameters in the field or comparing with numerical simulations. Since 2011, we have developed and improved out measurement method for this parameter and used this method throughout our prototype lasers' field tests. In this paper, we will introduce our measurement method and present a comparison between one of our latest field test results using this method and simulation result.

\section{Sites preparation and equipments setup}

As mentioned in equation \ref{eq:lidar equation}, several parameters has to be measured simultaneously to determine the coupling efficiency of the laser. These parameters and related measurement equipments used during our field tests are listed in table \ref{tbl: ce measurement equipments}. 

\begin{table}
	\begin{tabular}{|c|p{.3\linewidth}|p{.3\linewidth}|}
		\hline
		Parameter name & Measurement equipment & Description \\
		\hline
		\multirow{3}{.3\linewidth}{returned flux of laser guide star ($F$)} & photometry telescope & Planewave 12.5 inch Corrected Dall-Kirkham telescope \\\cline{2-3}
		& Johnson V band filter & Standard Johnson UBVRI filter \\\cline{2-3}
		& CCD camera & Princeton Instruments 512x512 electrical cooling camera\\
		\hline
		\minitab[c]{sodium column density ($C_{Na}$)\\ sodium layer centroid height ($L$)}& sodium Lidar & CSSC sodium Lidar\\
		\hline
		laser power ($P$) & power meter & ThorLabs PM100D with S120C sensor \\
		\hline
		atmosphere transmission ($T_a$) & auxiliary telescope & 25cm telescope\\
		\hline
	\end{tabular}
	\caption{Measurement equipments for measuring parameters in the Lidar equation } \label{tbl: ce measurement equipments}
\end{table}


\begin{figure}
\centering
\includegraphics[width=\linewidth]{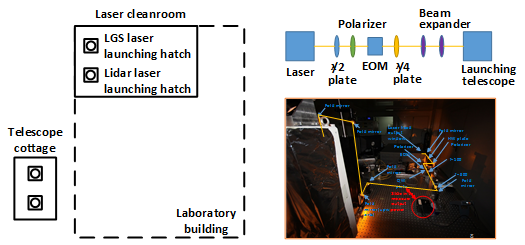}
\caption{(Left) Layout plan of the LGS coupling efficiency measurement facility for Xing Long test campaign. Similar layout was applied also for previous field tests. (Right upper) Layout plan of the beam transfer optics for the LGS laser. (Right lower) Actual layout of the laser bench. The red circle marks where asynchronous laser parameter measurement equipments can be switched in/out. } 
\label{fig:figure_1}
\end{figure}

The left pane of figure \ref{fig:figure_1} shows our setup used during 2015 Xing Long test campaign (\cite{FengLu2015IAU}, \cite{FengLu2016a}), similar layouts were adopted in our previous tests with minor changes to accomodate space constraints (Gao Mei Gu 2013 test, \cite{2014SPIE.9148E..3LJ}, Gao Mei Gu 2014 test, \cite{2015PASP..127..749J}, Canada Vancouver UBC 2013 test, \cite{Angel:2015}). The LGS laser and the sodium Lidar laser are located in a modified clean room in the laboratory building of the site. Two hatches are installed on the laboratory ceiling right above the Laser Launching Telescope (LLT) of the LGS laser and the zenith pointing fold mirror of the Lidar laser respectively for launching lasers to the zenith. A make-shift cottage is built 20 meters away from laser launching points. The choice of its location is limited by surrounding buildings and terrain, but decided not to be too far away from the launching points to minimize LGS spot elongation and synchronous delay for Lidar. The cottage also has two hatches installed on its roof. The 32cm LGS photometry telescope and a 50cm Dobsonian telescope for Lidar system are set up under these hatches respectively. A 25cm auxiliary telescope provided by Xing Long site is used routinely during test nights to monitor the atmospheric transmission.

The scheme for LGS laser beam transfer optics are shown in the right pane of \ref{fig:figure_1}. The 589.159nm sodium laser comes out from the port side of the package. A combination of the half-wave plate and the thin-film polarizer acts as power attenuator for adjusting projected laser power. An Electro Optics Modulator (EOM) is used to generate sodium D2$_b$ line sidebands from the original D2$_a$ line for D2$_b$ repumping technique which will bring enhancement for photon flux return (\cite{2009amos.confE..33K}). A Quarter-Wave Plate (QWP) is added after the EOM to adjust the polarization of the output laser beam. A pair of lens is inserted to adjust the laser beam width to fully fill the input aperture of the LLT. Large space is intentionally left between the 1st and 2nd fold mirrors after the beam expander. This space is for switching asynchronous parameter measurement setups, for instance, setups for measuring beam width, pointing, polarization, pulse shape, spectrum, et cetra. These measurements are done intermittently every night because most of these parameters are stable once settled. If one of these parameters is not stable, it can be observed by instability of wavelength or power, which are constantly monitored by a monitor stage inside the laser package. A tiny fraction of 589nm laser inside the laser package is guided to this stage. A wavelength monitor and a power meter are mounted and kept monitoring while the laser is on.

\section{Observation and data reduction}

\begin{figure}
	\centering
	\includegraphics[width=\linewidth]{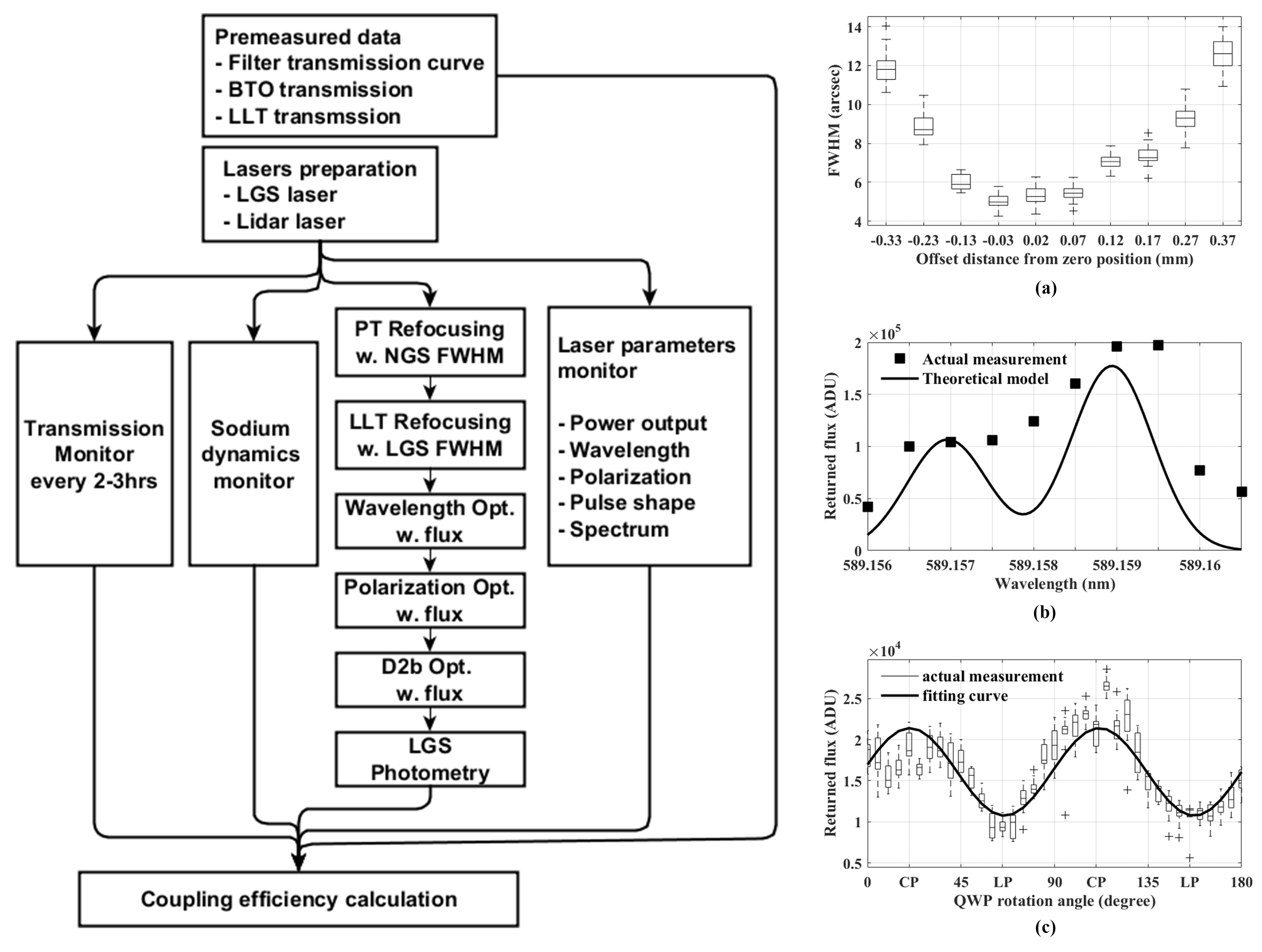}
	\caption{(Left) Laser guide star coupling efficiency measurement procedure. (Right) Parameter tuning before coupling efficiency measurement: (a) LLT focal stage tuning against LGS FWHM, (b) wavelenth tuning against LGS return flux, (c) polarization tuning against LGS return flux.}  
	\label{fig:figure_2}
\end{figure}

Observations for the laser guide star are only conducted in cloudless night. The procedure for laser guide star observation is decribed in figure \ref{fig:figure_2}. Before every observation, LGS laser has to be reoptimized to have the maximum and stable power output, as well as a stable wavelength near 589.159nm (D2$_a$ line). The LGS photometry telescope (PT) pointing to the zenith is refocused with natural stars. The LGS laser is then projected to the zenith. Because the laser beam has a gaussian profile, it is necessary to refocused the beam to the sodium layer by adjusting the focal length of the launching telescope (Figure 2(a)). The LGS return flux profile against wavelength is determined by the emission line of the sodium atom and the Doppler broadening mechanism. The profile shows that the highest return flux happens when the laser is tuned to the D2$_a$ line of the sodium atom(\cite{Steck}). The wavelength of the LGS laser is thus optimized by maximizing the return flux of the LGS (Figure 2(b)). It is reported by  (\cite{2009amos.confE..33K,Holzloehner2010}), that polarization of the laser beam could affect the pumping efficiency of the sodium laser, thereby affects the return flux of the LGS. The circular polarization has the highest return flux, while the linear polarization has the lowest. Although it is possible to measure the beam's polarization on the bench (right pane of figure \ref{fig:figure_1}), we choose to determine polarization by return flux because there are several reflecting surfaces after the test point that could alter the final polarization. The polarization optimization is done by adjusting the rotation angle of the QWP to where the return flux has the highest value (Figure 2(c)). By controlling the modulation depth of the EOM, the amount of power shifted from D2$_a$ to D2$_b$ is controlled. This fraction is optimized to around 10-15$\%$ of the total laser power (\cite{Holzloehner2010,Rampy2012,2015SPIE.9678E..1BF}). 

While the sodium laser guide star is in operation, we also keep monitoring the temporal variance of the sodium layer's column density $C_{Na}$ as well as its central height $L$ with the sodium Lidar. The atmospheric transmission $T_a$ is monitored every 2-3 hours by the 25cm telescope. 

We chose Johnson V-band filter because it is easy to acquire and relatively cheap. However, because the filter's spectrum range is much wider than the Doppler broadened spectrum width of the sodium atom, the application of this filter brings two disadvantages for photometry of the sodium laser guide star:
\begin{itemize}
	\item contamination from other wavelengths within the filter band reduces photometry precision,
	\item V magnitude of natural star is calculated by integrating all lights within the filter band because star's spectrum is wider than the filter bandwidth. Integrating LGS light within this band and compare with star's V magnitude will decrease the value of actual brightness of the LGS. 
\end{itemize} 
The first disadvantage is trivial for us because the magnitude of the laser guide star we generated is in the range of $v7$ to $v7.5$, sky background noise is trivial comparing to laser guide star's brightness. To solve the second problem introduced by the filter band mismatching with the narrow band of sodium fluorescence, we resort to natural reference star's spectrum rather than V magnitude value that were previsouly used for differential photometry. However, because the photometry telescope is pointing to the zenith, there is one more complication that during field tests we have to identify stars in LGS images that already have their spectra measured and logged in public accessible large spectral survey databases, such as LAMOST DR3 database (\cite{LAMOSTdr3}). 

The photometry of the LGS star is thus done by the following. The reference natural star and the laser guide star from image are extracted. The coordinates of the natural stars are then identified (\cite{1538-3881-139-5-1782}). We search stars' spectra $F(\lambda)$ by their coordinates in LAMOST DR3 database. If a star's spectrum could be found in the database, its spectrum along with its V magnitude will be used to calculate the LGS's photon flux. A normalized photon flux of the star in V band is then calculated with,
\begin{equation}
F_V = \frac{1}{hc}\int_{0}^{\infty}F(\lambda)W(\lambda)\lambda d \lambda
\end{equation}
where $W(\lambda)$ is the V band response function (multiplication of the V band filter transmission curve and Quantum Efficiency curve of the CCD). Since we know that at $\lambda_0=555.6nm$ for A0 star, Vega, the value of $F^{Vega}_{\lambda_0}$ is $9.4\times10^7 s^{-1}m^{-2}nm^{-1}$. Therefore, the normalized photon flux from the reference star, at $\lambda_0$, could be calculated by,
\begin{equation}
F_{\lambda}(\lambda_0) = 10^{-0.4V}\frac{F_{0V}^{Vega}}{F_{0V}}F_{\lambda}^{Vega}(\lambda_0)
\end{equation}
The absolute photon flux from the reference star at any wavelength $\lambda$ could then be found by,
\begin{equation}
F_{\lambda}(\lambda) = \frac{\lambda F_{\lambda}(\lambda)}{\lambda_0 F_{\lambda}(\lambda_0)}F_{\lambda}(\lambda_0)
\end{equation}
The instrumental flux measured by CCD, in ADC units, is given by,
\begin{equation}
F_{inst} = A\int_{0}^{\infty}F_{\lambda}W(\lambda)d\lambda
\label{eq:instrumental flux}
\end{equation}
where conversion factor $A$ is a proportionality constant related to the telescope area, throughput, CCD gain, and atmospheric transmission which does not vary in short time scale. Therefore, with an identified star in image, we could calculate $A$ by equation \ref{eq:instrumental flux} and apply this value in equation \ref{eq:lgs abs flux} to calculate the absolute photon flux of laser guide star.
\begin{equation}
F_{LGS} = \frac{F_{LGS inst}}{A W(589.159nm)} 
\label{eq:lgs abs flux}
\end{equation}
where $F_{LGS inst}$ is the laser guide star's instrumental flux in ADU unit measured from CCD. The coupling efficiency can be calculated by equation \ref{eq:lidar equation} after $F_{LGS}$ is determined.

\begin{figure}
	\centering
	\begin{subfigure}{.5\linewidth}
		\centering
		\includegraphics[width=\linewidth]{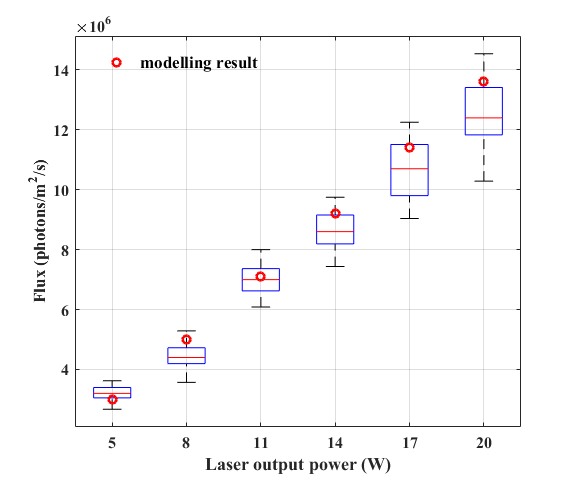}
		\caption{flux}
		\label{fig:comparsion-flux}
	\end{subfigure}%
	\begin{subfigure}{.5\linewidth}
		\centering
		\includegraphics[width=\linewidth]{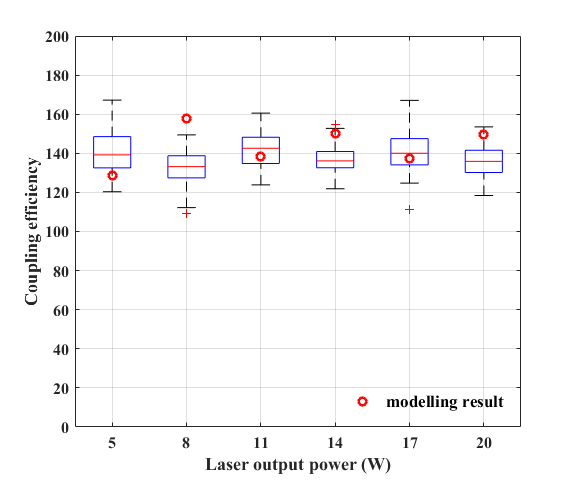}
		\caption{coupling efficiency}
		\label{fig:comparsion-ce}
	\end{subfigure}		
	\caption{Comparison between measurement results and modelling results. Each box in the boxplot contains 30 images of the actual measurements. }
	\label{fig:comparsion}
\end{figure}

In figure \ref{fig:comparsion}, we showed measured results of LGS flux and coupling efficiency measured with method described during one night of our 2015 Xinglong test campaign. The laser was working at 800Hz, circular polarization with 120ns pulse width and a 10\% power branching ratio for D2$_b$ repumping. A simulation was conducted with real-time measurment data such as laser spectrum, pulse profile, polarization state, laser direction, sodium column density as well as sodium layer height. Results from simulation is also plotted against the measurements which shows a good agreement between theoretical modelling and the measurement with this method.

\section{Conclusion}
In this paper, we showed a new method for measuring the performance of sodium laser in the generation of LGS. A new metric, the coupling efficiency of the laser which reflects the absolute performance for LGS generation, is used instead of the flux of LGS. A test setup for the measurement as well as the calculation method using Lamost spectrum data to calibrate filter effect is described. Coupling efficiency measurement results are compared with simulation results, and a good agreement is presented.

\normalem
\begin{acknowledgements}
This work was supported by the National Natural Science Foundation of China (11303056, 11273002), the NAOC astronomical financial special fund (Y533061V01), and Key Laboratory of Optical Astronomy, CAS. We acknowledge help from Xing Long and Gao Mei Gu Observatories. We greatly appreciate helps from Prof. Wang Ji Hong and Prof. Yang Guo Tao from National Space Science Center for their sodium Lidar support, Dr. Jia Ming Jiao from University of Science and Technology of China for his support in Lidar raw data analysis.
\end{acknowledgements}
  
\bibliographystyle{raa}
\bibliography{bibtex}

\end{document}